\documentclass{epl}
\usepackage{epsfig,psfrag}
\newcommand{\agt}{\mbox{\raisebox{-4pt}{$\,\buildrel>\over\sim\,$}}}

\title{Impurity effects in few-electron
quantum dots:\\  Incipient Wigner molecule regime}
\shorttitle{Impurity effects in few-electron
quantum dots} 
\author{B. Reusch \and R. Egger}
\institute{Institut f\"ur Theoretische Physik,
Heinrich-Heine-Universit\"at, D-40225 D\"usseldorf, Germany
}

\begin{document}

\maketitle

\begin{abstract}
Numerically exact path-integral
Monte Carlo data are presented for $N\leq 10$ strongly
interacting electrons confined in a 2D parabolic quantum dot, 
including a defect to break rotational symmetry.
Low densities are studied, where an incipient Wigner molecule forms.
A single impurity is found to cause drastic effects: (1) The standard
shell-filling sequence with magic numbers $N=4,6,9$,
corresponding to peaks in the addition energy $\Delta(N)$, is destroyed,
with a new peak at $N=8$, (2) spin gaps decrease,
 (3) for $N=8$,  sub-Hund's rule spin $S=0$ is induced, and (4)
spatial ordering of the electrons becomes rather sensitive to spin.
We also comment on the recently observed bunching phenomenon.
\end{abstract}

During the past few years, the electronic properties of
2D quantum dots have been the subject of intense theoretical studies
\cite{reimann,alhassid}.  A particularly interesting aspect of such
artificial atoms is the wide experimental tunability of the electron
number $N$ and the Brueckner interaction strength parameter $r_s$
in high-quality semiconductor dots \cite{ashoori96,kouwe01,ciorga}.
In particular, it
 has been established both theoretically \cite{reimann,egger} and
experimentally \cite{ashoori,bunching} that a ``Wigner molecule'' of
crystallized electrons forms already at
surprisingly high densities corresponding to $r_s\agt 2$ \cite{filinov}.
In this paper, we present numerically exact
path-integral Monte Carlo (PIMC)  \cite{suzuki}
simulation results for $N\leq 10$ electrons
in the most challenging incipient
Wigner molecule regime, choosing $r_s\approx 4$.
Due to the breakdown of effective single-particle descriptions,
this parameter regime is notoriously difficult to treat
within approximation schemes such as
Hartree-Fock theory \cite{reimann},  density functional
theory \cite{reimann,hirose1,hirose2},  semiclassical \cite{hausler}
or classical \cite{bedanov} approaches.
Especially concerning disordered dots, in the
absence of benchmark calculations,
their accuracy has so far been largely unclear.

The incipient Wigner molecule regime also exhibits
unexpected and interesting physics, in particular rather
dramatic effects when including just one
impurity in order to break rotational symmetry.
Below we shall demonstrate this in three different ways.
First, despite the presence of strong interactions,
the clean system still exhibits shell structure,
with exceptional stability of the dot for
``magic numbers'' $N=4,6$, and 9.  The stability is reflected
in peaks in the corresponding addition energy,
\begin{equation}\label{add}
\Delta(N)= E(N+1)-2E(N)+E(N-1),
\end{equation}
where $E(N)$ is the energy of the $N$-electron dot \cite{foot0}. 
Strikingly, a single impurity is able to completely alter this shell structure.
Only the filled-shell peak at $N=6$ survives, and
 a new peak at $N=8$ emerges. A similar  peak has also been
observed in recent experiments \cite{ciorga,kouwe01}.
A second example concerns the ground-state spin $S$,
where a high value of $S$ for certain $N$
can lead to a ``spin blockade'' for transport through
the dot \cite{weinmann}.
The role of impurities on the spin polarization of interacting quantum dots
has recently been studied by various groups, suggesting either a tendency
towards spin polarization \cite{berkovits}, depolarization \cite{stone},
or strongly $N$-dependent behaviors without general trend \cite{hirose2}.
As energy differences between different spin states are
typically very tiny, approximations are particularly prone
to predict false spin.
We find that ``spin gaps'' between the ground state and the next-higher spin
state are typically decreased by
the impurity, indicating a trend towards
 polarization, but without change in the ground-state
spin for the $N$ under study.
However, there is a notable exception to this rule:
For $N=8$, the rare case of sub-Hund's rule spin $S=0$ is found
in the presence of the impurity, while the clean dot has $S=1$.
Finally, as third example, we show that the electron crystallization pattern
shows a strong dependence
on total spin $S$, especially if rotational symmetry is broken.
To give a typical example, for $N=6$ electrons with
$S=0$, electrons basically arrange on a ring, but for
the fully polarized $S=3$ state, the amount of charge in the central region
increases by about $40\%$ compared to $S=0$.  Therefore, there is no clear
separation of energy scales for spin and charge fluctuations.
Such behavior is only found in the {\sl incipient}\
 Wigner molecule regime,
and invalidates the commonly used lattice-spin models appropriate in the
deep Wigner-crystallized limit \cite{hausler}.

\begin{table}
\caption{Quantum dot energies for fixed  $N$ and $S$ at $k_B T=0.125$
from PIMC simulations with and without impurity.}
\label{table1}
\begin{center}
\begin{tabular}{|cc|cc||cc|cc|}\hline
$N$ & $S$ &$E_{\rm imp}/\hbar\omega_0$&$E_{\rm clean}/\hbar\omega_0$&
$N$ & $S$ &$E_{\rm imp}/\hbar\omega_0$&$E_{\rm clean}/\hbar\omega_0$
\\\hline\hline
1 & 1/2 & 0.51(1)  & 1.00      &7 & 1/2 & 52.49(2) & 53.71(2)\\
2 & 0   & 3.911(7) & 4.893(7)  &7 & 3/2 & 52.555(23) & 53.80(2)\\
2 & 1   & 3.960(6) & 5.118(8)  &7 & 5/2 & 52.72(4) & 53.93(5) \\
3 & 1/2 & 9.857(8) & 11.055(8) &8 & 0   & 67.12(2) & 68.52(2) \\
3 & 3/2 & 9.880(8) & 11.050(10)&8 & 1   & 67.18(2) & 68.44(1)\\
4 & 1   & 17.89(2) & 19.104(6) &8 & 2 &  67.30(5)  & 68.51(5)\\
4 & 2   & 18.05(1) & 19.34(1)  &9 & 3/2 & 83.22(4) & 84.45(6) \\
5 & 1/2 & 27.75(1) & 29.01(2) & 9& 5/2   & 83.37(17)& 84.61(17)\\ 
5 & 3/2 & 27.84(2) & 29.12(2) & 10 & 1 & 100.59(11) & 101.96(14) \\
5 & 5/2 & 28.00(3) & 29.33(2) & & & & \\
6 & 0   & 39.30(2) & 40.53(1) & & & & \\
6 & 1   & 39.37(2) & 40.62(2) & & & & \\
6 & 2   & 39.48(2) & 40.69(2) & & & & \\
6 & 3   & 39.84(7) & 40.83(4) & & & & \\\hline
\end{tabular}
\end{center}
\end{table}

\begin{figure}
\centerline{\epsfxsize=8cm \epsfysize=7.5cm \epsffile{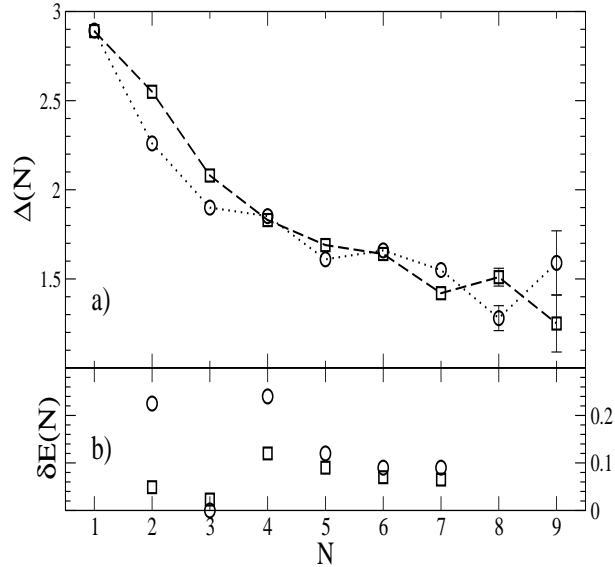}}
\caption{\label{fig1} PIMC results with impurity (squares) and for a clean dot
(circles), for a) the addition energy $\Delta(N)$ and b) the spin gap
$\delta E(N)$ for $N<8$. Error bars, unless shown explicitly, are of the order
of the symbol size. Dashed and dotted lines are guides to the eye only.
}
\end{figure}

We start by discussing the model underlying our work.
A closed 2D quantum dot with parabolic confinement of frequency $\omega_0$
and an additional impurity potential has been studied. Measuring energy
in units of $\hbar\omega_0$ and distance in oscillator
lengths $l_0=\sqrt{\hbar/m^*\omega_0}$,
the Hamiltonian for the $N$-electron system at zero magnetic field is
\begin{equation} \label{ham}
  H= \sum_{j=1}^N \left(\frac{1}{2}\,\bm{p}_j^2 +
   \frac{1}{2}\, \bm{r}_j^2 + V_{\rm imp}(\bm{r}_j) \right) +
   \sum_{i<j}^N \frac{\lambda}{|\bm{r}_i-\bm{r}_j|},
\end{equation}
where $\lambda=l_0/a$ for effective Bohr radius $a$.
Throughout this study, we set
 $\lambda=4$, which puts the quantum dot into
the incipient Wigner molecule regime with  $r_s\approx 4$ \cite{reimann}.
Here $r_s$ is half the nearest-neighbor distance of the electrons
measured in units of $a$.
For GaAs-based dots, this corresponds to $\hbar \omega_0\approx 0.7$~meV \cite{a
shoori,bunching}.
We allow for an attractive impurity
of strength $w$ centered at $\bm{x}$. The impurity potential is taken
in the form \cite{hirose2}
\begin{equation} \label{imp}
V_{\rm imp}(\bm{r})= -w \exp [ -2(\bm{r}-\bm{x})^2/\sigma^2 ] ,
\end{equation}
with parameters chosen as $w=4$,
$\bm{x} = (0,1.5)$, and impurity range $\sigma = 0.75$.
Qualitatively similar results were also observed for other
parameter choices.
For the above parameters, the impurity is close enough to
the center to influence the electrons. Typically,
when integrated over a circle of radius $\sigma$ around $\bm{x}$,
about $3/4$ of an electron charge is trapped by
this impurity. However, the impurity location is also
sufficiently far away from the center to break rotational symmetry.

Up to statistical error bars, PIMC data presented below are numerically exact.
For $\lambda=4$ and $k_B T=0.125$, the fermion
sign problem \cite{suzuki} permits us to extract reliable results only
for $N\leq 10$.  The simulations yield
 accurate estimates, at given $N$ and
total spin $S=(N_\uparrow-N_\downarrow)/2$, for the energy, the charge or spin
density, and other correlation functions of interest.
Typically, $5\times 10^7$ MC samples were accumulated
for each parameter set.  Errors due to finite
Trotter time discretization $\Delta \tau$ were carefully eliminated by
extrapolating results for different discretization down to
$\Delta\tau\to 0$ with a linear regression fit,
 using the fact that the expectation value of any Hermitian observable
has finite-$\Delta\tau$ corrections vanishing $\propto \Delta\tau^2$
\cite{fye}. This extrapolation procedure is essential for achieving
high accuracy, as Trotter convergence is very slow except
for fully spin-polarized clean dots \cite{foot1}.
PIMC results for the spin-dependent energy are listed in table~\ref{table1}, 
where bracketed numbers
denote one standard deviation error bar for the last digit,
which includes stochastic MC errors and extrapolation uncertainties.
When comparing our data in table~\ref{table1} to $T=0$ results
obtained from other methods such
as exact diagonalization \cite{merkt,mikha02,mikha02b},  one has
to take into account finite temperature, which can cause
sizeable differences for not fully spin-polarized states. 
For instance, for two electrons and $S=0$, PIMC predicts 
$E_{\rm clean}=4.893(7)$, while the $T=0$ result is $E=4.848$.  
Such differences arise because PIMC proceeds at fixed total $z$-component 
of the spin and therefore receives contributions from higher spin states 
 \cite{egger}.  Note that this effect is
rather tiny and does neither affect the (finite-temperature)
 addition energies (\ref{add}) nor the prediction of ground-state spins.  

Let us first discuss the addition energies displayed in fig.~\ref{fig1}a).
Remarkably, in a clean dot, one can still observe
the standard shell-filling sequence also found experimentally in small vertical
dots \cite{kouwe01,tarucha}.
The filled-shell configuration $N=6$ and the half-filled
cases $N=4, 9$ are exceptionally stable.
The peaks in the addition energy at these magic numbers signal
that the Wigner molecule is not  yet fully established,
since the classical Wigner limit is characterized
by different and much less pronounced magic numbers
\cite{bedanov}.
By simultaneously monitoring the charge density (see below), however,
clear traces of spatial ordering can be observed, reflecting
the onset of Wigner crystallization.
{\sl The energetic shell structure is then destroyed
once the impurity is present},
see fig.~\ref{fig1}a).  Apart from a new peak at $N=8$,
the addition energies $\Delta(N)$ now show a monotonic decrease with $N$.
Only the filled-shell peak at $N=6$ is still visible, yet considerably 
diminished. 
We note that the peak at $N=2$ is always absent for this
interaction strength \cite{reimann,ciorga}.

\begin{figure}
\psfrag{a}{a)}
\psfrag{b}{b)}
\psfrag{c}{c)}
\psfrag{d}{d)}
\centerline{\epsfxsize=8cm \epsfysize=8cm \epsffile{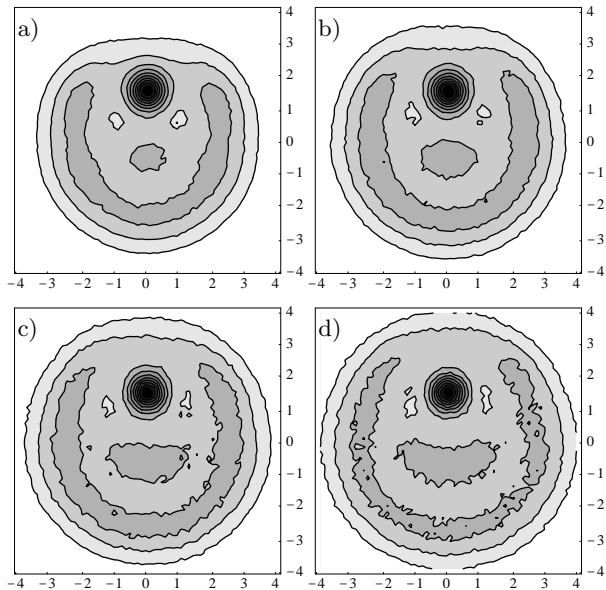}}
\caption{\label{fig2} Shadowed contour plots of the density
 (with impurity) for $N=7,8,9,$
and 10 [a)-d)].
Different contours lie at integral multiples of $0.1$ times
the maximal density; darker colors signal higher density.
In this figure, $k_B T=0.25$, with very similar but noisier results 
at lower $T$.}
\end{figure}

\begin{figure}
\psfrag{a}{a)}
\psfrag{b}{b)}
\psfrag{c}{c)}
\psfrag{d}{d)}
\centerline{\epsfxsize=8cm \epsfysize=8cm \epsffile{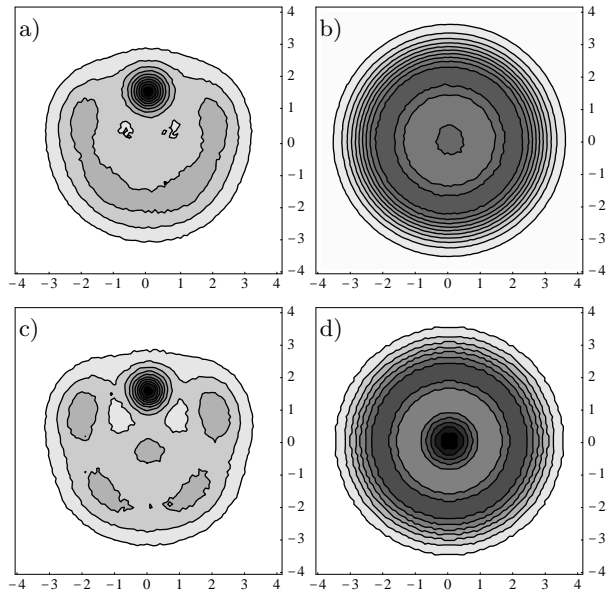}}
\caption{\label{fig3} Contour plots of the density for $N=6$.
a) With impurity, $S=0$. b) Clean case, $S=0$. c) With impurity,
$S=3$. d) Clean case, $S=3$.
}
\end{figure}

Next we address spin properties in this regime, see table~\ref{table1} and
fig.~\ref{fig1}b).  Except for $N=8$,
the spin $S$ was not altered by the impurity \cite{foot2}.
While $S$ mostly takes the minimal value ($S=0$ or $S=1/2$), for
$N=4$ $(N=9)$,  we have Hund's rule spin $S=1$ $(S=3/2)$. Furthermore, for
$N=10$, also a partially spin-polarized state $S=1$ is found.
For the clean $N=3$ dot, at $\lambda_c \approx 4.34$ a spin transition $S=1/2\to
 3/2$
occurs at $T=0$ \cite{egger,hausler,mikha02}.
We are already very close to this transition point,
and within error bars, we cannot decide the ground state spin.
The case $N=8$ is particularly noteworthy.  For the clean dot,
the spin is $S=1$ according to Hund's rule, but
the $S=2$ state is actually  very close,  see table~\ref{table1}.
Strikingly, with impurity, we then find
 {\sl sub-Hund's rule spin} $S=0$.
Since this effect requires sufficiently strong interactions (low density),
for a weakly disordered dot with $N=8$,
one can expect a transition from the usual
Hund's rule $S=1$ ground state at high density
(small $r_s$) to the unpolarized $S=0$ state at low density.
Since for $N=9$, we have spin $S=3/2$, this transition could be
observed experimentally as spin blockade for the 9th electron
to enter the dot.  By monitoring spin-spin correlations between the
electrons at the impurity location and the central region
(data not shown here),
strong antiferromagnetic couplings are visible.  Hence it is likely that local
moment formation at the impurity location and in the central region is
important for this sub-Hund's rule behavior.  This is also confirmed by noting
that charge and spin densities do not follow each other,
as they do in the deep Wigner-molecule regime \cite{egger,mikha02b}.
Since except for $N=8$ the spin $S$ is not changed,
the impurity's overall effect on the spin polarization
seems small.
Nevertheless, our results show a decreasing ``spin gap''
$\delta E(N)= E(S+1)-E(S)$,
see fig.~\ref{fig1}b), indicating a weak tendency towards polarization.
Note that the spin excitation energies are very small,
again stressing the need for highly accurate calculations.

\begin{figure}
\centerline{\epsfxsize=6cm \epsfysize=4.5cm \epsffile{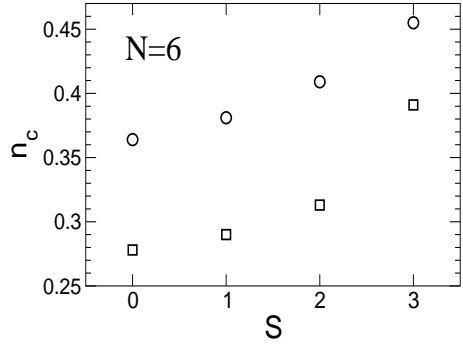}}
\caption{\label{fig4} Amount $n_c$ of electron charge localized in
the center (see text) for $N=6$ and different $S$, both with (squares)
and without (circles) the defect.
}
\end{figure}

We now turn to the onset of electron crystallization
in real space. Figure \ref{fig2} shows density plots
for $N=7$ to 10 electrons in their respective ground-state spin configuration.
At the defect, the density is about three times larger than
elsewhere in the dot.
Although there are large quantum fluctuations, a spatial shell structure
is already discernible. The classical prediction for clean dots
at these electron numbers is as follows \cite{bedanov}.
For $N=7$ and $N=8$, just one electron is in the center, while
for $N=9$ and $N=10$, the spatial filling is $(2-7)$ and $(2-8)$,
respectively, with the configurations $(1-8)$ and $(3-7)$ very close in energy.
Quantum-mechanically, for $\lambda=4$, the radial ordering is not
yet very pronounced in the clean dot \cite{egger,reima00}. With impurity,
however, our data in fig.~\ref{fig2} displays spatial shells, although with
a rather different filling sequence. Detailed examination of the central
region shows that only one to two electrons are accommodated in this
part. Therefore incoming electrons enter the outer ring for $N=7,8,9,$
and 10, with the ring expanding and, for $N=10$, nearly embracing the
impurity.
While for very large $r_s$, the
crystal structure is $S$-independent and effective Heisenberg-type lattice
models for the spin physics can be constructed,  for $r_s\approx 4$,
the {\sl spatial charge ordering is strongly dependent on total spin} $S$;
see also~\cite{mikha02b, reima00}.
This is seen in fig.~\ref{fig3} for $N=6$, where the
densities for $S=0$ and $S=3$ are
compared, both with and without the defect.  Interestingly,
for $S=0$, electrons tend to basically arrange on a ring, while for the
spin-polarized case, the classical configuration
with one electron in the center \cite{bedanov,reima00}
is approached.  This effect, in
particular concerning the influence of the impurity, can be
 quantified by computing the fraction $n_c$ of electron charge
localized within a radius $\sigma=0.75$ around the center of the
dot.  The result is shown in fig.~\ref{fig4}, and illustrates that
more and more charge goes into the central region when $S$
is increased.  For the clean dot, comparing $S=0$ and $S=3$, this
amounts to a $\approx 25\%$ increase in $n_c$, but with the defect,
this increase is $\approx 40\%$.  That implies that the spin sensitivity
of the incipient Wigner crystallization process is
enhanced by the impurity, probably due to a local spin
at the impurity site.

Finally, let us briefly discuss the relation of our results
to the experimentally observed ``bunching'' phenomenon
found in the Wigner molecule regime \cite{bunching,ashoori98},
where the addition energy was found to vanish at  certain values of $N$.
Bunching was also reported in a recent numerical
diagonalization study for $N=7$ and $N=8$ \cite{canali},
where the dot was modeled by a $3\times 4$
Hubbard-type model containing only short-range interactions.
In contrast to these results, for
the more realistic continuum Hamiltonian (\ref{ham})
with long-range interactions, bunching is {\sl not}\ observed for $N < 10$,
see fig.~\ref{fig1}a).
Since bunching is believed to be a generic effect \cite{ashoori98},
a consistent explanation
probably has to assume strong disorder or other mechanisms
not contained in our model.

To conclude, we have presented a PIMC study of
impurity effects in 2D quantum dots.
 We have focused on the most difficult
yet most interesting regime of an incipient Wigner molecule. Large effects
are caused already by a single impurity.  When the rotational symmetry of the
clean dot is broken, magic numbers for the energetic shell-filling
sequence, observed as peaks in the addition energy, are changed.
Instead of the usual peaks at $N=4,6,9$, with impurity only a (weak)
filled-shell peak at $N=6$ and a new peak at $N=8$  are observed.
In addition, for $N=8$, we find a sub-Hund's rule spin $S=0$.
Finally, the incipient Wigner crystallization process
was found to be strongly spin-dependent, in particular
when the impurity is present. We hope that these results
will also be valuable as reference data to assess the validity
of computationally less demanding yet approximate calculations.

\acknowledgments
We thank H. Grabert, W. H{\"a}usler, and T. Heinzel for discussions.
This work has been supported by the DFG under the Gerhard-Hess program.


\begin{thebibliography}{0}
\bibitem{foot0}
{Experimentally observable addition
energies are determined by {\sl free energy} differences. However, 
at the low temperature considered here, entropic contributions
are very small.  This has been checked for several $N$.}
\bibitem{foot1}
{Further improvement is possible using the
multi-level blocking approach \cite{egger} to alleviate the sign problem. 
However, it was not used here to allow for arbitrary $\Delta \tau$.}
\bibitem{foot2}
{For $N=9,10$, the  ground-state spins were derived from 
finite-$\Delta \tau$ data not given in table~\ref{table1}.}
\end{thebibliography}

\end{document}